# Quantum Paraelectric Glass State in $SrCu_3Ti_4O_{12}$

Jitender Kumar, Ram Janay Choudhary, and A.M. Awasthi*
UGC-DAE Consortium for Scientific Research, University Campus, Khandwa Road, Indore- 452 001, INDIA

## Abstract

Magnetic and dielectric studies of $SrCu_3Ti_4O_{12}$ carried out over 5-300K confirm antiferromagnetic (AFM) ordering of Cu-spins at $T_N$ =23K. Dielectric constant $\varepsilon'$ measured across 1Hz-1MHz signifies quantum paraelectric character, Barrett-fittable almost down to $T_N$. Competition of athermal fluctuations and the literature-reported magneto-phonon-softening near $T_N$ manifests a quantum paraelectric glass (QPG) state. Emergent AFM-field tunes the otherwise quantum ordering (at absolute-zero) of the dipoles to finite-temperature kinetic glass transition; spectral dispersion of dielectric constant unambiguously manifested and characterized. Vogel-Fulcher glass-kinetics parameterization sets the almost relaxation-free QPG state in $SrCu_3Ti_4O_{12}$ apart from an emergent scaling-class, to which typical ferroelectric relaxors belong.

---

The physics of quantum fluctuations and its effects on the properties of the host-materials have recently gained prominence and attention of condensed matter researchers. Resultant emergent phenomena/states are fascinating and important from the viewpoint of basic and materials science; two of them being quantum paraelectrics[1] (QP) and quantum spin-liquids.[2] Quantum paraelectrics are important due to their huge piezoelectric effect at cryogenic temperatures,[3] whereas quantum spin-liquids play dominant role in the low-temperature metal-insulator transitions, Mott-insulators, and superconductivity.[2] Prime effect of quantum fluctuations is to prevent a macroscopic ordering down to 0K, by sustaining an athermal disordered state.[4] In recent years, excellent articles have appeared on superconductivity of the hybrids of graphene and a magnetic material, wherein due to the proximity effects, a superconducting state is observed at the nano-scale.[5] According to Loffe and Michael,[6] dynamical inhomogeneity due to quantum fluctuations hinders the long range superconductivity in graphene. A general schematic of fig.1 represents the systems with quantum fluctuations having spin/dipolar degrees of freedom. The possible emergent matter-states include (*i*) magneto-electric multiferroics[7] from the coupling of long-range-ordered magnetic and electrical degrees of freedom, (*ii*) quantum paraelectrics[1] from athermal fluctuations of dipolar degrees of freedom, (*iii*) quantum spin liquids[2] from athermal fluctuations of magnetic spins, and (*iv*) disordered states from the interplay of coupled spin and dipolar degrees of freedom and athermal fluctuations; quantum paraelectric glass (QPG), quantum spin glass (QSG), or a quantum multi-glass (QMG). Of these, QSG would be akin to a 'multiferroic' Griffith's phase, driven by a long-range electrical order. For example, the emergent (magneto-electric) multiglass state[8] in (Sr,Mn)$TiO_3$ is realized, though by static disorder (Mn-doping) in the parent $SrTiO_3$ quantum paraelectric.

* Corresponding Author e-mail: amawasthi@csr.res.in. Tel: +91 731 2463913.

In quantum paraelectrics (QP's), the ferroelectric (FE) ground state is suppressed (no FE-$T_C$ down to 0K) by the quantum zero-point fluctuations;[1] they are also called the *incipient ferroelectrics*.[1,9] The prototype quantum paraelectrics[1,10-11] are $SrTiO_3$, $KTaO_3$, and $CaTiO_3$. QP's show huge and $T$-independent dielectric constant at cryogenic temperatures.[1] This is realized due to the subtle balance of soft-phonon mode and quantum fluctuations; therefore, non-thermal external influences like electric field, pressure, or impurity-doping can create a ferroelectric or a relaxor ground state.[10,12-13] In nearly all of the QP's, a quantum critical phase transition[4] (QPT) results from external-tuning of the (otherwise) 0K transition to finite temperatures; e.g., a ferroelectric or relaxor state is achieved in non-magnetic (Ba, Bi)-doped $SrTiO_3$,[14-15] Pb-doped $CaTiO_3$,[16] and in Li-doped $KTaO_3$.[17] The disorder created by the doping of a magnetic atom is also quite able to suppress the quantum fluctuations of dipole-moments, resulting in the polar nano-regions (PNR's) and/or spin-glass behavior.[9] Another important quantum paraelectric $EuTiO_3$ shows AFM transition at the Nèel temperature $T_N$ =5.5K,[18] below which the dielectric constant drops. $EuTiO_3$ shows magneto-dielectric effect near $T_N$, since Eu-spins are strongly coupled with the soft-phonon mode.

We present $SrCu_3Ti_4O_{12}$ (SCTO) as the latest quantum paraelectric, which undergoes AFM-driven electrical vitrification, resulting from the competition of athermal quantum dipolar fluctuations and the relevant-phonon-softening accompanying the G-type antiferromagnetic (AFM) ordering[19] at 23K. Here, the emergent magnetic field of the ordered spin-system tunes the otherwise 0K transition (as per indicated by the Barrett-fit permittivity) to the observed finite-temperature quantum critical kinetic phase transition, realizing the QPG state. Thus, SCTO is second to $EuTiO_3$ in deviating at finite-$T$ under internal field, from its parent QP character. SCTO belongs to the $ACu_3Ti_4O_{12}$ family of copper-titanates; better known as colossal dielectric constant (CDC) materials, viz., $CaCu_3Ti_4O_{12}$ (CCTO).[20] Both SCTO and CCTO have cubic double-Perovskite structure with space group $Im3$,[19-21] but have huge differences in their dielectric properties. Moreover, the low-temperature dielectric investigation of SCTO (down to liquid-Helium range) and the prospects of magneto-dielectricity have not been explored, which we present here.

The ceramic SCTO samples were prepared from high purity (99.99%) powders of $SrCO_3$, CuO, and $TiO_2$ by the conventional solid state route. For making of good quality samples we ground the mixed charge of precursors for more than 45 hrs. and calcined it at 1050°C for 24 hours. The pelletized samples (10 mm diameter and 1-3mm thick) were sintered at 1100°C for 24 hours and their flat faces were silver-coated to make good electrical contacts for the dielectric measurements. X-Ray diffraction of the samples has been done with Cu-$K_\alpha$ radiation ($\lambda$ =1.54Å), using a Bruker D8 Advance X-ray Rotating-anode powder diffractometer. Dielectric measurements over 4.2K-room temperature spanning 1Hz to 1MHz with 1V ac-excitation were performed using (Alpha-A) High Performance Frequency Analyzer (NOVO-CONTROL). The magnetization data were collected from 2-300K using 7-Tesla SQUID-vibrating sample magnetometer (SVSM; Quantum Design Inc., USA).

Phase-purity and crystal structure of the samples were analyzed by the Rietveld analysis using the fullprof software. Rietveld refinement ($\chi^2$ ~ 2.2, goodness of fit 1.5) provided the lattice constant 7.4050(1)Å for the cubic space group $Im3$ without any secondary phase. The crystal structure (VESTA software) of SCTO is shown in fig.2, made using the fitting parameters obtained from the Rietveld refinement. Like other family members,[20,22] SCTO too has tilted Ti-$O_6$ octahedra (direction $x$=0→1, $y$=0.25→0.75, $z$=0.25→0.5), as shown in fig.2 inserts. This octahedral-tilt makes the Ti-O-Ti bond-angle 141.83(14)° (as per determined from our results, resembling an earlier report[19,21]) instead of 180°, and forms square-planar arrangement of Cu-$O_4$, with Cu at the center and O's at the corners.[20] Quantum paraelectricity in SCTO is rooted in its crystal structure. For example, $BaTiO_3$ is ferroelectric while (Sr/Ca)$TiO_3$ are quantum paraelectrics. Ionic-size of Ba is larger than that of Sr/Ca, thus providing larger space for the Ti-$O_6$ octahedral cage to expand. Therefore, the $Ti^{+4}$-ions rattle easily in the former and as a result $BaTiO_3$ undergoes series of phase transitions, which is not possible in (Sr/Ca)$TiO_3$.[23] $SrCu_3Ti_4O_{12}$ (SCTO) also has the tilted Ti-$O_6$ octahedral structure and the Ti-O bond-length 1.958(4)Å is similar to that in $SrTiO_3$ (1.95Å); i.e., smaller compared to 2Å in $BaTiO_3$. Suppressed rattling/displacement of $Ti^{+4}$ in the octahedral-cage kills bulk ferroelectricity in the centric SCTO, while fluctuating short-range dipolar correlations impart the system quantum paraelectricity.

Temperature dependent magnetization $M(T)$ of SCTO at 100 Oe is shown in fig.3, resembling an earlier published report.[19] The AFM order is observed at the Nèel temperature $T_N$ =23±0.01K, close to that reported for CCTO,[24] with little observable difference in ZFC and FC data. High magnetic field up to 7T does not affect the $T_N$, confirming rather robust exchange interaction responsible for the G-type AFM order. By Curie-Weiss linear-fit of the $1/\chi$-$T$ data, $\Theta_{C\text{-}W}$ = -39.1±0.1K and effective magnetic moment $\mu_{eff}$/Cu-ion = 2.09±0.0006$\mu_B$ are evaluated (fig.3, right $y$-axis). The metric of magnetic frustration[25] $f = |\Theta_{C\text{-}W}|/T_N$ ~1.73 for SCTO is though larger than $f \approx 1$ for CCTO; implying presence (absence) of AFM fluctuations above their respective $T_N$'s in SCTO (CCTO), also indicated by the deviation of the SCTO $1/\chi$-$T$ data from the perfect Curie-Weiss fit at $T \geq T_N$ (fig.3 inset). In SCTO, only Cu ($Cu^{2+}$, $d^9$) carrying $s$ =1/2 spin in the $3d$ shell orders collinearly along the crystallographic [111] direction.[19] The first- and third- nearest neighbors interact antiferromagnetically, whereas the interaction between the second-neighbor Cu-ions is ferromagnetic in nature. A direct interaction between $Cu^{2+}$ ions is scarce, because the distance between these ions is quite large. Indirect super-exchange between Cu-ions is mediated through the $Ti^{4+}$ ions (similar to CCTO),[24] endowing SCTO more direct magneto-dielectricity; $Ti^{+4}$ cations being also the constituent of electric-polarizability in the Ti-$O_6$ octahedra provide a platform for major influence of magnetic ordering on the electrical degrees of freedom.

Figure 4 shows the dielectric constant of SCTO from 4.5K to room temperature at 800 kHz, the high-frequency most clearly providing both the classical and quantum temperature-regimes as seen below. Sample quality plays an important role in the dielectric characterization of SCTO; e.g., the dielectric constant of impure samples is higher in comparison to the pure one.[19] In the case of our specimen, the value of dielectric constant (~73) is close to the intrinsic value predicted by the first principles theory (for similarly-structured CCTO, which should be less than 100 at room temperature[26]), ensuring very good sample quality, free from any static/structural disorders.[19] It is therefore suggested that the extrinsic (Maxwell-Wagner) effects present in CCTO (responsible for its huge dielectric constant) are absent in SCTO[21] over similar $T$-range. Almost down to $T_N$, the temperature dependence of real permittivity fits well the Barrett formula,[27] the theoretical model available for the quantum paraelectrics, given as

$$\varepsilon'(T) = A + C/\left[(T_1/2)\coth(T_1/2T) - T_C\right],$$

here $T_C$= -63.35±3.9K is equivalent to $\Theta_{C\text{-}W}$ = -419±0.7K, obtained from the classical (high-$T$) Curie-Weiss linear-fit. Below ~$T_1$, the two (quantum and classical) behaviors are supposed to split; happening at ≈155K, independent of the probed frequencies. Barrett-fit to the data at 800kHz provides $T_1$ =154.67±2.5K, indicating that the high-temperature classical behavior is accurately reflected in the Barrett parameters derived from high frequency data. In the inset, we show the dramatic rise of the normalized deviance $\left[(\varepsilon'_B/\varepsilon'_{CW}) - 1\right]$ below $T_1$, as a metric of the net quantum paraelectric (QP) character. Barrett's turnover to plateau makes this QP-metric drop below ~50K. Barrett fit confers QP character to SCTO; high-$T$ antiferroelectric correlations of dipoles ($\varepsilon'$-$\varepsilon'_{CW}$ split below $T_1$ and -ve value of $T_C$) exclude their low-$T$ organization into *polar* nano-regions (PNR's of relaxors), or to a robust ferroelectric state.

The dipolar and spin degrees of freedom in SCTO are directly coupled as follows. Individual oxygens of $Ti\text{-}O_6$ octahedron are each bonded to a different Cu-atom and each $Cu\text{-}O_4$ forms a square-planar arrangement;[20] and this sharing tilts the $Ti\text{-}O_6$ octahedra, making the Ti-O-Ti bond-angle 141.83° instead of 180°.[19,21] As these oxygens are associated with the Ti-O phonon, their sharing between $Ti\text{-}O_6$ octahedra and $Cu\text{-}O_4$ square-planes (reflected in the octahedral-tilt) means a change in Cu-spin arrangement affects the Ti-O phonon. This spin-phonon coupling determines the magneto-electricity in SCTO, as also reported e.g., in $DyMn_2O_5$.[28] Raman signatures of spin-phonon coupling in SCTO have been recently reported;[29] below the Nèel temperature $T_N$, Cu-spins arrange antiferromagnetically and the associated Ti-O phonon ($A_g(1)$ rotation-like mode at 442cm$^{-1}$) softens, registering the observed drop (fig.4) in the dielectric constant. When the system undergoes AFM transition, the strong internal magnetic field tends to induce a long-range electrical ordering due to the spin-phonon coupling. On the other hand, the athermal QP-fluctuations oppose this tendency; the compromise being the medium-range dipolar-organization into nano-scale clusters. In the related $Na_{1/2}Bi_{1/2}Cu_3Ti_4O_{12}$, recently Ferrarelli *et al*.[30] using infrared/THz spectroscopy have reported incipient-ferroelectric/quantum-paraelectric character. The soft-mode frequency was fitted by a modified Barrett formula, to determine the relevant temperatures $T_1$ and $T_C$; however, no effect of the AFM-ordering on the dielectric behavior was reported. Their dielectric data taken at GHz range (for intrinsic part sans huge Maxwell-Wagner contributions) is similar to ours' presented here on SCTO.

With reference to fig.5a, the magnetic correlations onset their kinetic (frequency-dependent) effect of decreasing the dielectric constant (otherwise undergoing eventual Barrett level-off) almost 10K above $T_N$. While this *dynamic manifestation* of the clusters is triggered by the fluctuating AFM-correlations existing *above* $T_N$ ($f$ = $|\Theta_{C\text{-}W}|/T_N$ > 1), their *static manifestation* is the continued $\omega$- dependence of $\varepsilon'$ *below*, due to the underlying quantum fluctuations (QF). Without these QF's, the magnetic-frustration alone may cause a magneto-dielectric $\varepsilon'$-dispersion *strictly above* $T_N$; with its sub-$T_N$ demise (i.e., $\varepsilon'(\omega)$-merger sans QF) at all frequencies, contrary to our results as per obtained. Clear low-$T$ frequency-dispersion in the dielectric constant of a spatially-uniform (sans defects/doping) QP-parent is observed here, reflecting nano-scale electrical-segmentation dynamically (statically) above (below) $T_N$. We attribute it to the magneto-electric competition product of the internal magnetic field, coupled with the quantum-fluctuating dipole-moments, and thermal energy. It is important to note that while in the classical FE-relaxors, the *static disorder* inhibits the long-range electrical ordering, the *dynamical disorder* here is caused by the athermal quantum fluctuations.[6] Pure SCTO QP-parent thus becomes the first to feature vitreous dispersive-response character of a kinetic phase transition; the observed state is qualified to be coined as ***Quantum Paraelectric Glass*** **(QPG)**.

Dispersion-kinetics of dielectric constant here confirms the Vogel-Fulcher-Tammann (VFT) glassy-slowdown[31-32] (fig.5a, inset) of the characteristic frequency $\omega_p(T) = \omega_0\exp[-E_a/(T-T_0)]$, which generally describes dispersion in the *relaxation frequency* of FE-relaxors. Note that the Arrhenius ($\omega_p$-$T^{-1}$) plot here also serves the purpose of a thermo-spectral 'dynamic phase-boundary'; separating the 'liquidus & glassy' regimes under & above the same, of the QP degrees of freedom. The curve marks upper (lower) cut-off frequency (temperature) at a particular temperature (frequency), for the dielectric and piezoelectric response of the liquidus (unarrested) phase to be manifested, probed, and manipulated. Moreover, the observed high-$T$ classical paraelectricity (sans dipolar-correlations) marks $T_1 \approx 155$K as the upper cut-off temperature for this liquidus regime. Definitely, an electric-field ($\boldsymbol{E}_{dc}$) should nucleate & stabilize electrically-ordered regions in this liquidus regime (i.e., electrical 'solidification') besides altering the phase-boundary (i.e., narrowing the $\omega_p$-dispersion). Thus in SCTO too, electro-mechanical response[33] ought to be observable under bias-field, across $T_N$=23K to $T_1 \approx 155$K and up to excitation frequency $\omega_p(T,\boldsymbol{E})$, in the piezoresponse force microscopy (PFM). For the SCTO-QPG, VFT temperature from the fit is $T_0$ =13K and the activation energy is $E_a$ =25meV.

Over the range where the dielectric constant clearly displays measurable $\omega$-$T$ dispersion; fig.5b shows low-magnitude loss-background $\varepsilon''(T)$ without a peak-structure. This is understandable for two reasons. Firstly, for the antiparallely-clustered dipoles, configurations directed along or opposite to an applied $\boldsymbol{E}$-field are energetically equivalent. Therefore, the two local minima of their configurational potential-energy are degenerate; symmetric double-well rendering 'relaxations' mute for the *non-polar* nano-clusters, under the removal/flipping of the applied field. Secondly, at least above $T_N$, the AFM-correlations induce only *dynamic/transient* non-polar clusters, whose relaxation is meaningless. As such, $\omega_p(T)$ obtained from the dispersed $\varepsilon'(\omega,T)$-peaks here signifies characteristic *response-frequency* of these *non-polar* nano-clusters. In contrast, the same obtained for FE-relaxors (from either $\varepsilon'(\omega,T)$- or $\varepsilon''(\omega,T)$-peaks) refers to the *relaxation*-frequency of their *polar* nano-regions (PNR's). Therefore, for applications, QPG's seem more suitable than FE-relaxors, in that they provide a broadband (in both $\omega$ and $T$) high dielectric susceptibility, against the background of weakly-dispersive *marginal-losses*.

Dielectric losses ($\varepsilon''$) being negligent here (fig.5b), the QPG contrasts with the FE-relaxors, as revealed by the very distinct values for the two benchmarks used-in/describing their VFT behaviors. Generally, the ratio $T_0/T_g$ (limit 0 to 1) of ultimate ($\omega_p(T_0) \sim 0$) to ambient freezing temperatures ($T_p$ at 1kHz probing frequency, say) measures the *non-Arrhenicity* of dispersion-kinetics, whereas $E_a/k_BT_0$ (VFT-temperature-scaled barrier-activation energy) known as the *glass-strength*[34] is a metric of the resistance against devitrification of the glassy state by external means (c.f., pressure $\boldsymbol{P}$[35] for the structural glasses and electric field $\boldsymbol{E}$[36] for FE-relaxors). Table I compiles these metrices for a number of classical (statically-disordered) FE-relaxors, along with the same for the present (dynamically-disordered) SCTO-QPG, and those characterizing the glassy domain-wall freezing in $KH_2PO_4$ (KDP) crystal.[37] Apart from the qualitatively obvious[34] reverse-regression between these tabulated parameters *across* the types of relaxor/glassy specimens, we find that the family of FE-relaxors defines an exclusive scaling to which SCTO does not belong (fig.5b, inset). Therefore, clearly distinct anti-regressions between non-Arrhenicity and glass-strength delineate the categories of FE-relaxors and QPG. Moreover, the much-larger glass-strength for our SCTO-QPG translates into its feeble susceptibility to *electrical-devitrification*, which characterizes electrically-glassy FE-relaxors. Well-known *electrical crystallization* of the FE-relaxors under high $\boldsymbol{E_{dc}}$-fields into robust ferroelectrics[36] is thus little expected for the pure SCTO, from its glassy-regime (i.e., below ~ $T_N$). In retrospect, this also explains why the internal magnetic field due to the long-range AFM-ordering too fails to induce a bulk electrical order in SCTO, expected of a non-local magneto-electric coupling, and rather settles for the nano-scale electrically-vitrified state.

**Table I.** A Compilation of glass-kinetics parameters in arrested electrical degrees of freedom.

| **Materials**[Ref] | **Non-Arrhenicity** ($\boldsymbol{T_0/T_g}$) | **Glass-Strength** ($\boldsymbol{E_a/k_BT_0}$) |
|---|---|---|
| **$\mathbf{SrCu_3Ti_4O_{12}}$ (QPG)**[present] | **0.482** | **22.1** |
| 0.22BS-0.25PMN-0.53PT[38] | 0.916 | 1.91 |
| 0.9PMN-0.1PT[39] | 0.922 | 1.63 |
| PZN[40-41] | 0.942 | 1.37 |
| $KH_2PO_4$ (Domain-Wall Freezing)[37] | 0.957 | 0.20 |
| 0.75PMN-0.25PT[42] | 0.969 | 0.51 |

A functional-interest of the QPG state is the character of its electrical quality factor, defined as the inverse loss-tangent, $Q = \text{Cot}\delta = \varepsilon'/\varepsilon''$. Large value and spectrally/thermally benevolent behavior (enabling calibrations) of this $Q$-metric is practically important and desirable for the use of a dielectric in microwave/high-frequency device components such as resonators, oscillators, phase-shifters, and mixers for narrow-band applications. A major benefit of replacing the air-filled metallic-voids/cavities etc. by a dielectric is the size-downscaling of particulate structures by the refractive index $n = \sqrt{\varepsilon}$, crucial for miniaturization and large-scale integration. Moreover, due to the lower thermal expansion vs. metals, size-specific precision spectral parameters (e.g., operational frequency) of the device-structures incorporating the dielectrics remain sturdier against thermal variations. To this end, fig.6 shows this quality-factor spectrum for the SCTO-QPG, at key temperatures across the observed $\varepsilon'$-dispersion range. Note the rather high ~$O(10^3)$ magnitude (±4%) and regular $(\omega,T)$-variation (-10dB drop over 10Hz-1MHz and +7dB increase across 20-35K) functional-features of this $Q$-factor for SCTO-QPG. Remarkable too is the positive temperature-coefficient ($dQ/dT > 0$) of the quality factor.

To conclude, we have observed a quantum paraelectric glass (QPG) state in pure $SrCu_3Ti_4O_{12}$. The high-temperature QP-liquid state in competition with the AFM-order-driven phonon-softening ($A_g(1)$ rotation-like mode) is witnessed to undergo kinetic glass-phase transition near $T_N$=23K. Strong spin-phonon coupling due to the $Ti^{4+}$ cations, common to both the indirect Cu-Ti-Cu exchange and the Ti-O bond-polarizability, together with the frustrated magnetic correlations above $T_N$ results in peculiar magneto-dielectricity of this material. High-temperature antiferroelectric-like correlations intrinsic to the QP-parent exclude any polar-organization of the dipoles at low-temperatures. Essentially capacitive, low-loss magneto-dielectric response around $T_N$ features glassy Vogel-Fulcher frequency-dispersion (electrical-vitreousity), traced to non-polar nano-clusters. An interesting classification scheme sets the QPG state in SCTO distinctly apart as a strong electrical-glass-former, less susceptible to long-range electrical ordering under an $\boldsymbol{E}$-field, versus the FE-relaxors defining a family of *fragile glass-formers*. Rather small (≤ 5%) peak-anomaly in the dielectric constant near $T_N$ despite a direct (Ti-mediated) magneto-dielectric coupling is attributable to this "ideal-glass-like" and "non-polar" characters of the QPG state here; reflecting only the polarizability-change due to the electrical nano-clustering, without their polar-alignment/conglomeration (typical of the FE-relaxors). However, the nearly negligent-losses ($\tan\delta \sim 10^{-3}$) and the temperate $(\omega,T)$-dependent high-$Q$ ($=\varepsilon'/\varepsilon''$) characteristics of the QPG state ensure even this small (magneto-dielectric) $\Delta\varepsilon'$ to be robust/integral against time and other (electrical, magnetic, mechanical etc.) disturbances, meriting QPG preferable to the "polar-base" multiferroics, for prospective applications.

**Acknowledgements:** Authors thank Mukul Gupta for providing the X-Ray Diffraction data and Pankaj Pandey for help with the magnetization measurements.

## References:

[1] K.A. Müller and H. Burkard, Phys. Rev. B **19**, 3593 (1979).
[2] A.C. Potter, M. Barkeshli, J. McGreevy, and T. Senthil, Phys. Rev. Lett. **109**, 077205 (2012).
[3] D.E. Grupp and A.M. Goldman, Science **276**, 392 (1997).
[4] M. Vojta, Rep. Prog. Phys. **66**, 2069 (2003).
[5] A. Allain, Z. Han, and V. Bouchiat, nature materials **11**, 590 (2012).
[6] L.B. Ioffe and M.E. Gershenson, nature materials **11**, 567 (2012).
[7] W. Eerenstein, N.D. Mathur, and J.F. Scott, Nature **442**, 759 (2006).
[8] V.V. Shvartsman, S. Bedanta, P. Borisov, W. Kleemann, A. Tkach, and P.M. Vilarinho, Phys. Rev. Lett. **101**, 165704 (2008).
[9] S.N. Gadekar, *Proceedings of Nuclear and Solid State Physics Symposium*, Bangalore, India, p229 (1973).

[10] W.R. Abel, Phys. Rev. B **4**, 2696 (1971).

[11] V.V. Lemanov, A.V. Sotnikov, E.P. Smirnova, M. Weihnacht, and R. Kunze, Solid State Commun. **110**, 611 (1999).

[12] J. Hemberger, P. Lunkenheimer, R. Viana, R. Bohmer and A. Loidl, Phys. Rev. B **52**, 13159 (1995); S. Horiuchi, Y. Okimoto, R. Kumai, and Y. Tokura, Science **299**, 229 (2003).

[13] J.G. Bednorz and K.A. Muller, Phys. Rev. Lett. **52**, 2889 (1998).

[14] V.V. Lemanov, E.P. Smirnova, P.P. Syrnikov, and E.A. Tarakanov, Phy. Rev. B **54**, 3151 (1996).

[15] C. Ang, Z.Yu, P.M. Vilarinho, and J.L. Baptista, Phy. Rev. B **57**, 7403 (1998).

[16] V.V. Lemanov, A.V. Sotnikov, E.P. Smirnova, and M. Weihnacht, Appl. Phys. Lett. **8**1, 886 (2002).

[17] H. Yokota and Y. Uesu, J. Korean Physical Society **51**, 744 (2006).

[18] T. Katsufuji and H. Takagi, Phys. Rev. B **64**, 054415 (2001).

[19] D. Mori, M. Shimoi, Y. Kato, T. Katsumata, K.-I. Hiraki, T. Takahashi, and Y. Inaguma, Ferroelectrics **414**, 180 (2011).

[20] M.A. Subramanian, D. Li, N. Duan, B.A. Reisner, and A.W. Sleight, J. Solid State Chem. **151**, 323 (2000).

[21] J. Li, M.A. Subramanian, H.D. Rosenfeld, C.Y. Jones, B.H. Toby, and A.W. Sleight, Chem. Mater. **16**, 5223 (2004).

[22] J.-C. Zheng, A.I. Frenkel, L. Wu, J. Hanson, W. Ku, E.S. Božin, S.J.L. Billinge, and Y. Zhu, Phys. Rev. B **81**, 144203 (2010).

[23] R.E. Newnham, *Structure-Property Relations* (Springer-Verlag, New York, 1975).

[24] A. Koitzsch, G. Blumberg, A. Gozar, B. Dennis, A.P. Ramirez, S. Trebst, and Shuichi Wakimoto, Phys. Rev. B **65**, 052406 (2002).

[25] A.P. Raminez in *Handbook of Magnetic Materials*, edited by K.H.J. Buschow (North–Holland, Amsterdam, 2001), Vol. **13**.

[26] L. He, J.B. Neaton, M.H Cohen, D. Vanderbilt, and C.C. Homes, Phys. Rev. B **65**, 214112 (2002).

[27] J.H. Barrett, Phys. Rev. **86**, 118 (1952).

[28] N. Hour, S. Park, P.A. Sharma, S. Guha, and S.-W. Cheong, Phys. Rev. Lett. **93**, 107207 (2004).

[29] D.K. Mishra and V.G. Sathe, J. Phys.: Condens. Matter **24**, 252202 (2012).

[30] M.C. Ferrarelli, D. Nuzhnyy, D.C. Sinclair, and S. Kamba, Phys. Rev. B **81**, 224112 (2010).

[31] H. Vogel, Z. Phys. **22**, 245 (1921).

[32] G.S. Fulcher, J. Am. Ceram. Soc. **8**, 339 (1925).

[33] R.K. Vasudevan, D.M. Marincel, S. Jesse, Y. Kim, A. Kumar, S.V. Kalinin, and S. Trailer-McKinstry, Adv. Func. Mats. **23**, 2490 (2013); A. Kumar, Y. Ehara, A. Wada, H. Funakubo, F. Griggio, S. Trailer-McKinstry, S. Jesse, and S.V. Kalinin, J. Appl. Phys. **112**, 052021 (2012).

[34] *Dosorder Effects on Relaxational Processes*, edited by R. Richert and A. Blumen (Springer-Verlag, Berlin, 1994).

[35] E. Williams and C.A. Angell, J. Phys. Chem. **81**, 232 (1977).

[36] G.A. Smolinskii, V.A. Bokov, V.A. Isupov, N.N. Krainik, R.E. Pasynkov, and A.I. Sokolov, *Ferroelectrics and Related Materials* (Gordon and Breach, New York, 1981).

[37] J. Kumar and A.M. Awasthi, Appl. Phys. Lett. **103,** 132903 (2013).

[38] C.J. Stringer, T.R. Shrout, and C.A. Randall, J. Appl. Phys. **101**, 054107 (2007).

[39] D. Viehland, S.J. Jang, L.E. Cross, and M. Wuttig, J. Appl. Phys. **68**, 2916 (1990).

[40] G. Burns and F. Dacol, Phys. Rev. B **28**, 2527 (1983).

[41] C. Tu and C. Tsai, J. Appl. Phys. **87**, 2327 (2000).

[42] D. Vielhand, "*The Glassy Behavior of Relaxor Ferroelectrics*," Ph.D. thesis, The Pennsylvania State University (1991).

**Figure Captions:**

**Fig.1.** Schematic of the possible emergent matter-states from the interaction of spin and dipolar degrees of freedom, in the presence of their quantum fluctuations. QPG (quantum paraelectric glass), QSG (quantum spin glass), and QMG (quantum multiglass).

**Fig.2.** Rietveld-refined X-ray diffraction pattern of $SrCu_3Ti_4O_{12}$, along with its crystal structure in the direction ($x$=0→1, $y$=0.25→0.75, $z$=0.25→0.5), with tilted Ti-$O_6$ octahedra. Right sketch shows the coplanar arrangements of oxygens around the Cu, responsible for the octahedral-tilting.

**Fig.3.** Magnetization $M(T)$ of $SrCu_3Ti_4O_{12}$ with antiferromagnetic phase transition at $T_N$ =23K in little-different FC and ZFC runs (left $y$-axis). Inverse susceptibility $1/\chi$ vs. temperature (right $y$-axis) and the Curie-Weiss fit ($\Theta_{C\text{-}W}$ = -39.1K, $\mu_{eff}$/Cu-ion =2.09$\mu_B$, also enlarged in the right-inset). Top inset compares zoomed-in $M/H$ at 100 Oe and 7T, confirming no change in $T_N$ at high-fields, indicative of rather robust exchange interaction. Symbol-sizes represent the raw-data uncertainty.

**Fig.4.** Left $y$-axis: dielectric constant of $SrCu_3Ti_4O_{12}$ vs. temperature at 800 kHz, with Barrett fit (quantum paraelectric behavior) and showing the drop near $T_N$. Right $y$-axis: inverse electrical-susceptibility $(\varepsilon'\text{-}1)^{-1}$ vs. temperature; Curie-Weiss straight line and the Barrett fits split below $T \approx$ 155K (the quantum regime) and merge within uncertainty at higher temperatures (classical regime). Symbol-size represents the raw-data uncertainty. Inset: a normalized metric $\left[\left(\varepsilon'_{B}/\varepsilon'_{CW}\right)-1\right]$ of the net quantum paraelectric (QP) character, rising sharply at ~155K, maximizing at ~50K, and dipping at lower-$T$'s (reflecting Barrett's turnover to plateau-behavior).

**Fig.5.** (a) Glassy dispersion of the dielectric constant over a wide range (~ 6 decades) of frequency. Inset: Arrhenius-plot of probing frequency vs. inverse of $\varepsilon'$-peak temperature ($1/T_p$) fits the Vogel-Fulcher glassy slowdown with cooling. Regions under (above) the curve represent liquidus-QPL (glassy-QPG) dynamical regimes of the quantum paraelectric degrees of freedom. (b) Low-valued/featureless losses ($\varepsilon'' \sim 10^{-1}$ at $T_p$'s) signify no "polar-like" relaxations. Symbol-sizes represent the raw-data uncertainty. Inset: a 'phase' diagram of correlated glass-metrices (strength-against-devitrification vs. non-Arrhenicity of glassy kinetics) brings out the apartheid of QPG-SCTO presented here (open star) and the classical FE-relaxors (open dots, defining a clear family-locus), along with the data representing the domain-wall freezing (solid dot).

**Fig.6.** Spectral character at key temperatures of the quality factor $Q = \varepsilon'/\varepsilon''$ illustrates its regular -10dB drop over five decades in frequency and +7dB increase across 20-35K, for the SCTO-QPG. Highly-desirable positive temperature-coefficient of $Q$ is in contrast to its negative values for the metal-based structures, used in the high-frequency applications.

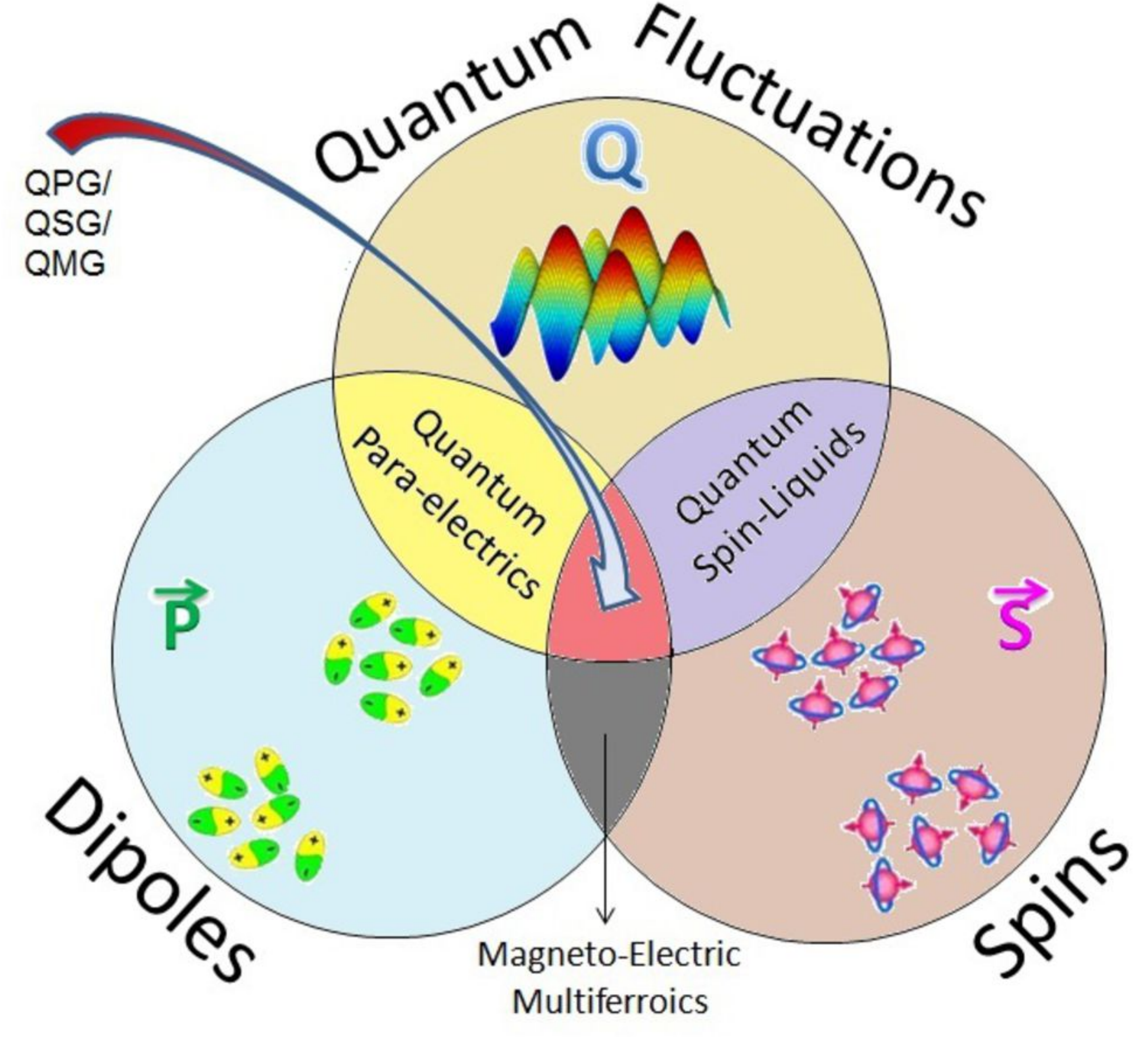

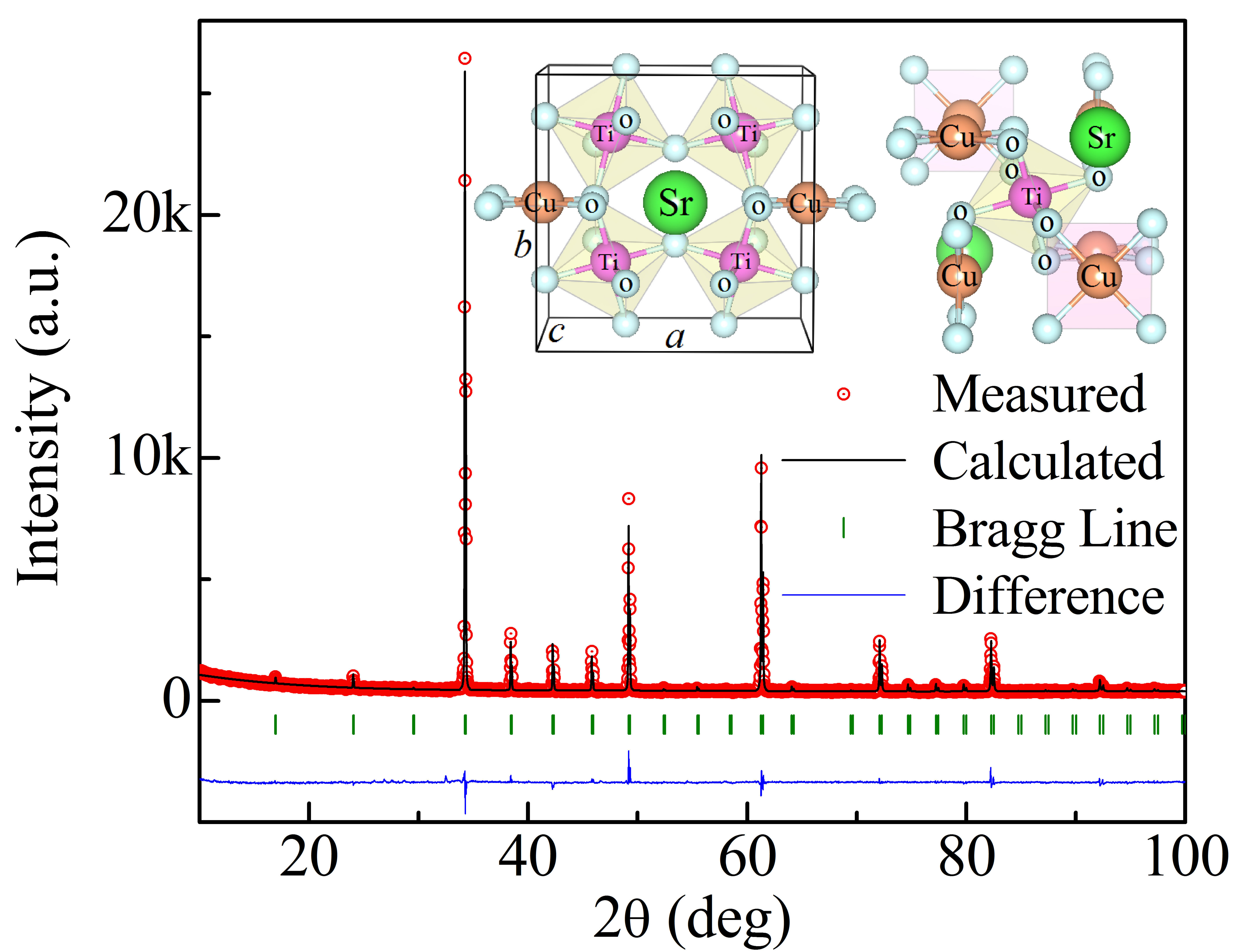

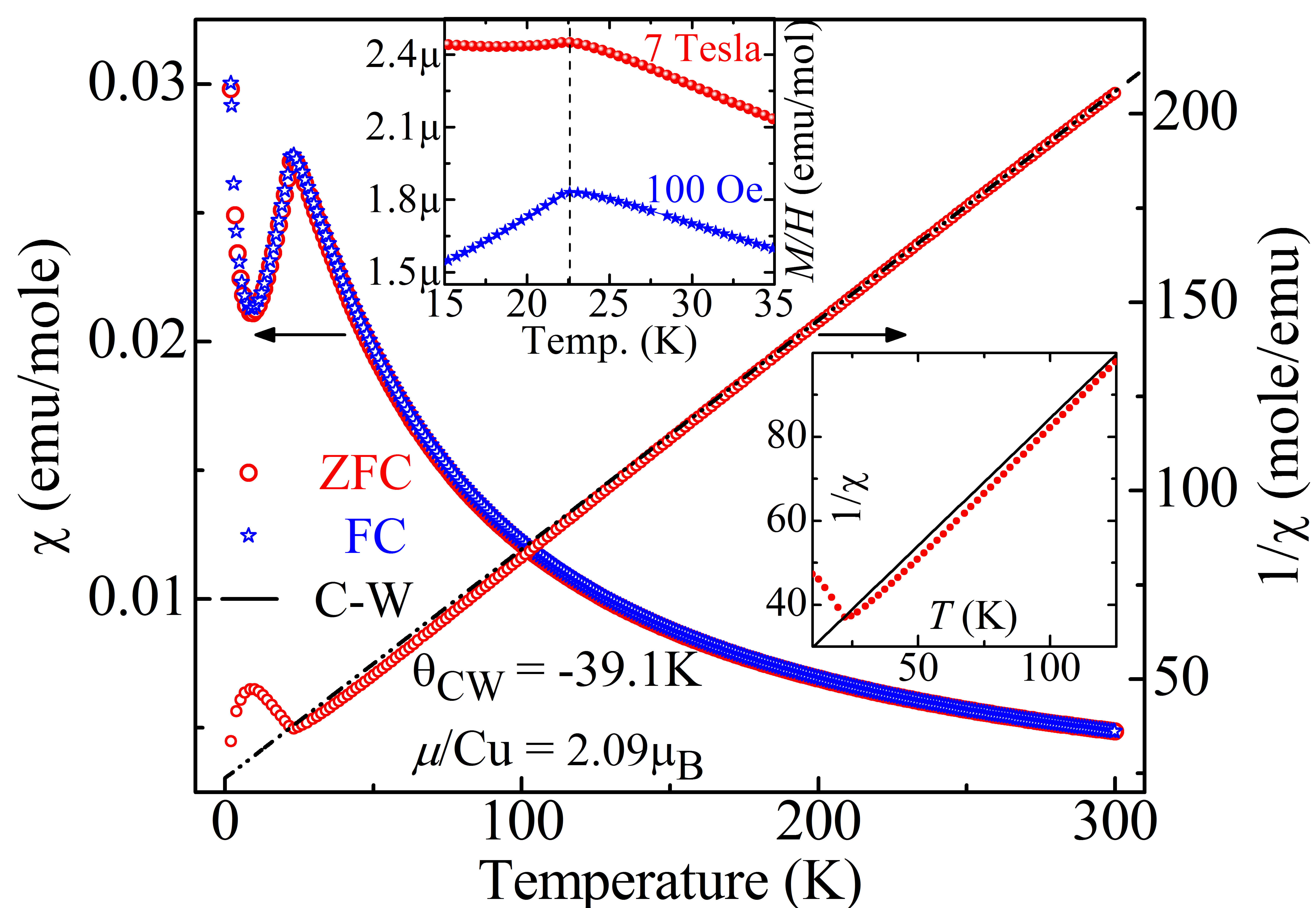

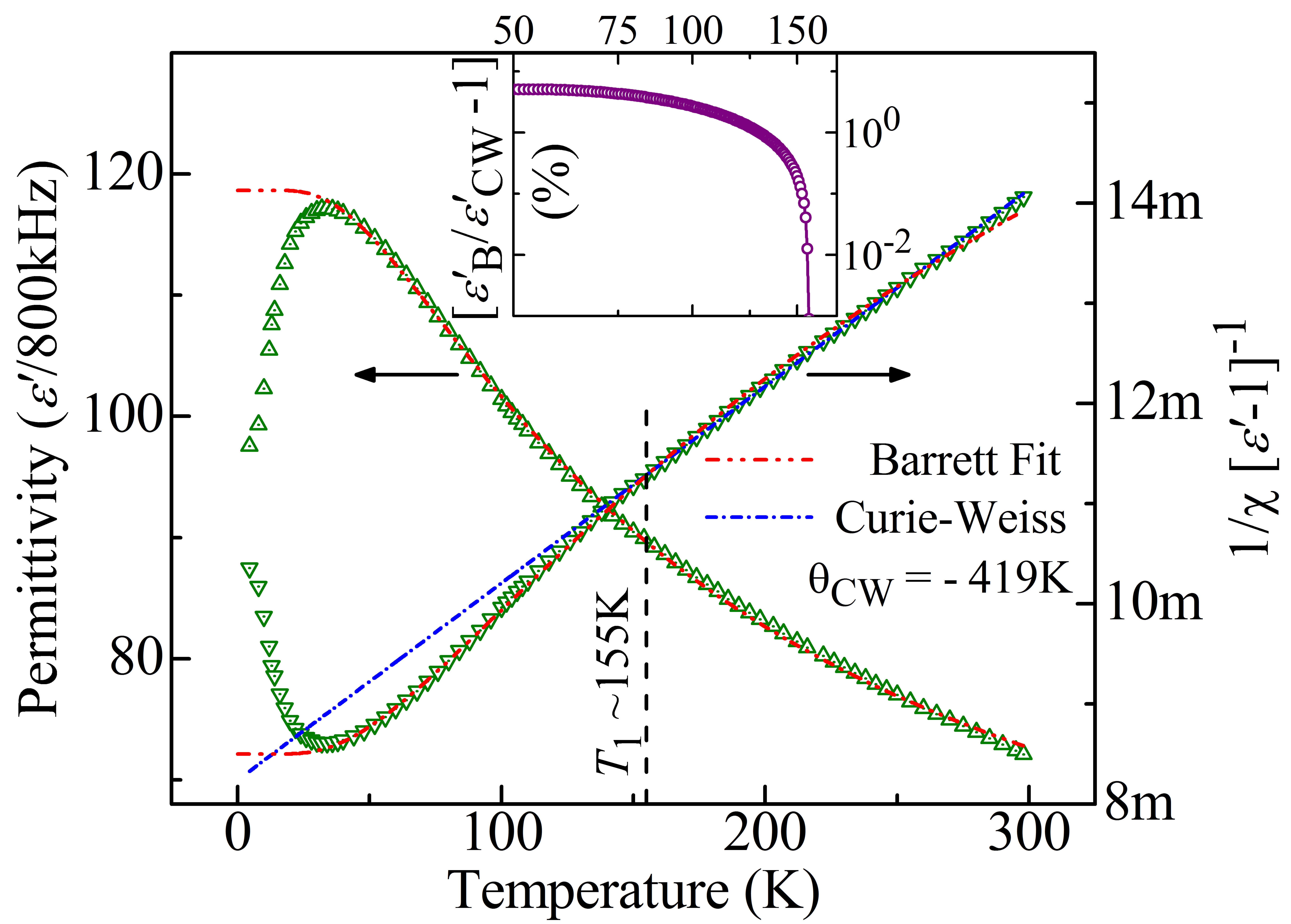

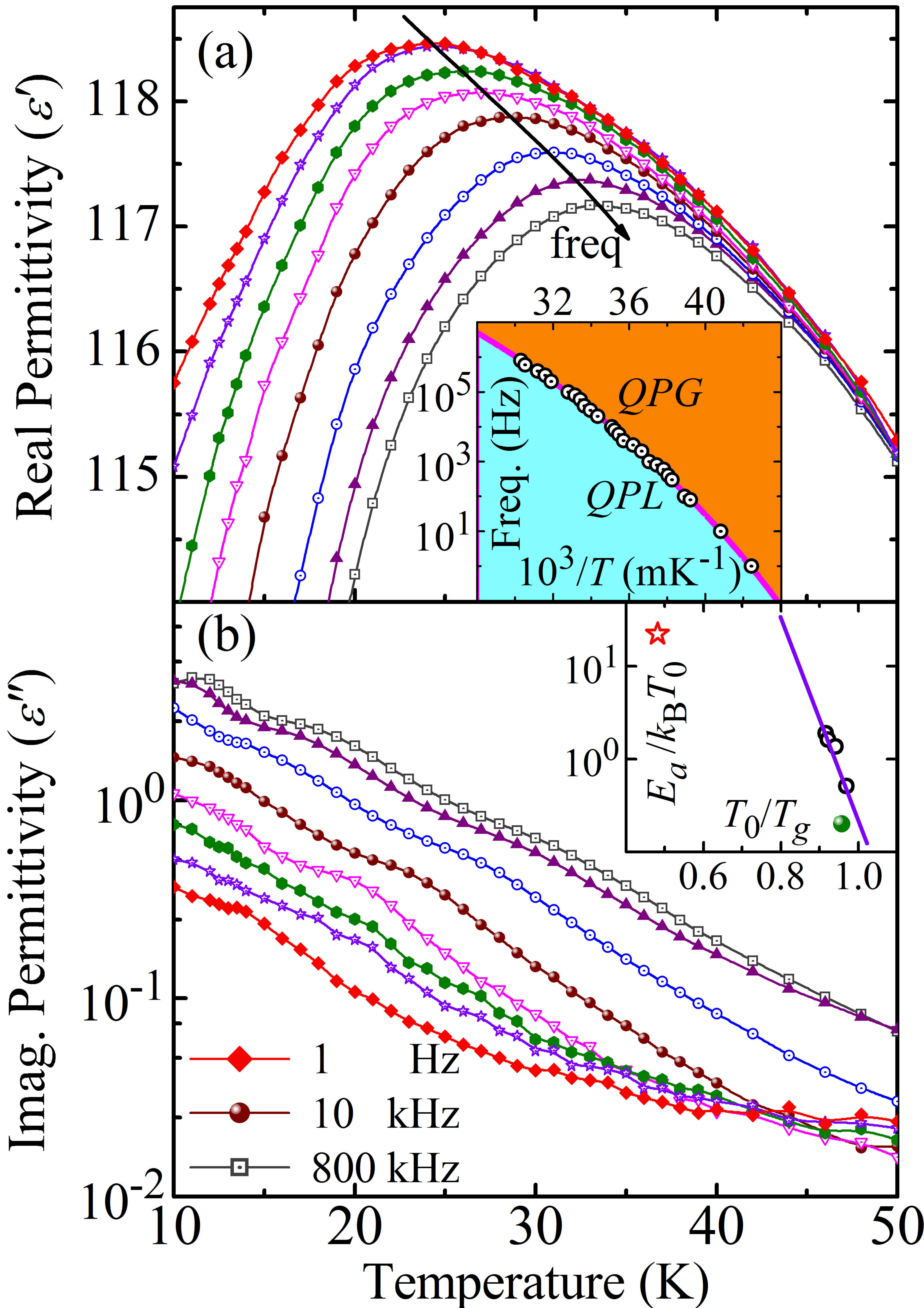

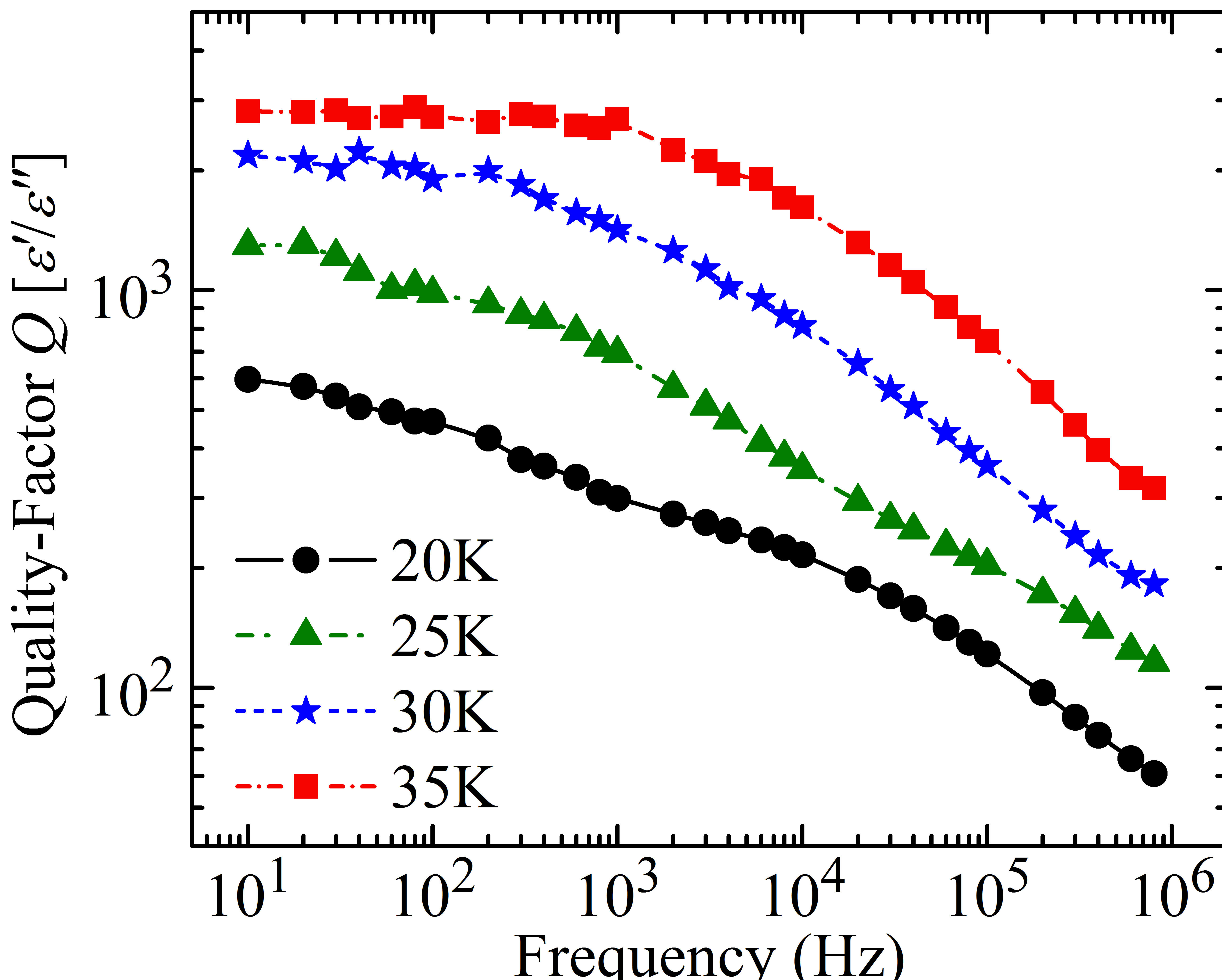